%% file: straub-moriond.tex
\documentclass{moriond}
\usepackage{amsmath}
\usepackage{graphicx}

\newcommand{\Photo}{}

\begin{document}
\title{\boldmath Implications of $b\to s$ measurements~\footnote{Talk presented at the 50th Rencontres de Moriond (Electroweak Session), La Thuile, 20 March 2015.}}

\author{  Wolfgang Altmannshofer$^1$ and David M. Straub$^{2,}$\,\footnote{Speaker.}}

\address{$^1$ Perimeter Institute for Theoretical Physics, 31 Caroline St. N, Waterloo, Ontario, Canada N2L~2Y5}

\address{$^2$ Excellence Cluster Universe, TU M\"unchen, Boltzmannstr.~2, 85748~Garching, Germany}

\maketitle\abstracts{The recent updated angular analysis of the $B\to K^*\mu^+\mu^-$ decay by the LHCb collaboration is interpreted by performing a global fit to all relevant measurements probing the flavour-changing neutral current $b\to s\mu^+\mu^-$ transition. A significant tension with Standard Model expectations is found. A solution with new physics modifying the Wilson coefficient $C_9$ is preferred over the Standard Model by $3.7\sigma$. The tension even increases to $4.2\sigma$ including also $b\to se^+e^-$ measurements and assuming new physics to affect the muonic modes only. Other new physics benchmarks are discussed as well.
The $q^2$ dependence of the shift in $C_9$ is suggested as a means to identify the origin of the tension -- new physics or an unexpectedly large hadronic effect.
}

\section{Introduction}

Rare $B$ and $B_s$ decays based on the $b\to s$ flavour-changing neutral current transition are sensitive to physics beyond the Standard Model (SM). Recent measurements at the LHC, complementing earlier $B$-factory results, have hugely increased the available experimental information on these decays. Interestingly, several tensions with SM predictions have shown up in the data, most notably
\begin{itemize}
\item several tensions at the $2$--$3\sigma$ level in $B\to K^*\mu^+\mu^-$ angular observables in 1~fb$^{-1}$ of LHCb data taken during 2011~\cite{Aaij:2013qta};
\item a $2.6\sigma$ deviation from lepton flavour universality (LFU) in $B^+\to K^+\ell^+\ell^-$ decays measured by LHCb, including  the full 3~fb$^{-1}$ dataset~\cite{Aaij:2014ora}.
\end{itemize}
Several model-independent theoretical analyses~\cite{Descotes-Genon:2013wba,Altmannshofer:2013foa,Beaujean:2013soa,Hurth:2013ssa,Alonso:2014csa,Hiller:2014yaa,Ghosh:2014awa,Hurth:2014vma} have shown that both anomalies could be explained by new physics (NP).
Today, the LHCb collaboration has released an update of the analyis of $B\to K^*\mu^+\mu^-$ angular observables based on the full 3~fb$^{-1}$ dataset~\cite{LHCb-NEW}, finding a significant tension in particular in the angular observable $P_5'$.
The aim of this talk is to interpret these measurements by performing a global model-independent fit to all available data. The results are updates of an analysis published recently~\cite{Altmannshofer:2014rta} (and building on earlier work~\cite{Altmannshofer:2011gn,Altmannshofer:2012az,Altmannshofer:2013foa}), incorporating the new LHCb measurements. Crucially, the fit makes use of a combined fit~\cite{Straub:2015ica} to $B\to K^*$ form factors from light-cone sum rules~\cite{Straub:2015ica} and lattice QCD~\cite{Horgan:2013pva,Horgan:2015vla} published recently.

\section{Model-independent analysis}

\subsection{Fit methodology}

The effective Hamiltonian for $b\to s$ transitions can be written as
\begin{equation}
\label{eq:Heff}
{\cal H}_{\text{eff}} = - \frac{4\,G_F}{\sqrt{2}} V_{tb}V_{ts}^* \frac{e^2}{16\pi^2}
\sum_i
(C_i O_i + C'_i O'_i) + \text{h.c.}
\end{equation}
Considering NP effects in the following set of dimension-6 operators,
\begin{align}
O_7 &= \frac{m_b}{e}
(\bar{s} \sigma_{\mu \nu} P_{R} b) F^{\mu \nu},
&
O_7^{\prime} &= \frac{m_b}{e}
(\bar{s} \sigma_{\mu \nu} P_{L} b) F^{\mu \nu},
\\
O_9 &= 
(\bar{s} \gamma_{\mu} P_{L} b)(\bar{\ell} \gamma^\mu \ell)\,,
&
O_9^{\prime} &= 
(\bar{s} \gamma_{\mu} P_{R} b)(\bar{\ell} \gamma^\mu \ell)\,,\label{eq:O9}
\\
O_{10} &=
(\bar{s} \gamma_{\mu} P_{L} b)( \bar{\ell} \gamma^\mu \gamma_5 \ell)\,,
&
O_{10}^{\prime} &=
(\bar{s} \gamma_{\mu} P_{R} b)( \bar{\ell} \gamma^\mu \gamma_5 \ell)\,,\label{eq:O10}
\end{align}
one can construct a $\chi^2$ function which quantifies, for a given value of the Wilson coefficients, the compatibility of the hypothesis with the experimental data. It reads
\begin{equation}
\chi^2(\vec C^\text{NP})
=
\left[\vec O_\text{exp}-\vec O_\text{th}(\vec C^\text{NP})
\right]^T
\left[\mathcal C_\text{exp}+\mathcal C_\text{th}\right]^{-1}
\left[\vec O_\text{exp}-\vec O_\text{th}(\vec C^\text{NP})
\right].
\label{eq:chi2}
\end{equation}
where $O_\text{exp,th}$ and $\mathcal C_\text{exp,th}$ are the experimental and theoretical central values and covariance matrices, respectively. All dependence on NP is encoded in the NP contributions to the Wilson coefficients, $C^\text{NP}_i=C_i-C^\text{SM}_i$. The NP dependence of $\mathcal C_\text{th}$ is neglected, but all correlations between theoretical uncertainties are retained.
Including the theoretical error correlations and also the experimental ones, which have been provided for the new angular analysis by the LHCb collaboration, the fit is independent of the basis of observables chosen (e.g.\ $P_i'$ vs.\ $S_i$ observables). In other words, the ``optimization''~\cite{Descotes-Genon:2013vna} of observables is automatically built in.

In total, the $\chi^2$ used for the fit contains 88 measurements of 76 different observables by 6 experiments (see the original publication\cite{Altmannshofer:2013foa} for references). The observables include $B\to K^*\mu^+\mu^-$ angular observables and branching ratios as well as branching ratios of $B\to K\mu^+\mu^-$, $B\to X_s\mu^+\mu^-$, $B_s\to \phi\mu^+\mu^-$, $B\to K^*\gamma$, $B\to X_s\gamma$, and $B_s\to\mu^+\mu^-$.

\subsection{Compatibility of the SM with the data}

Setting the Wilson coefficients to their SM values, we find
$\chi^2_\text{SM}\equiv\chi^2(\vec 0) = 116.9$ for 88 measurements, corresponding to a $p$ value of $2.1\%$.
Including also $b\to se^+e^-$ observables\footnote{We have not yet included the recent measurement~\cite{Aaij:2015dea} of $B\to K^*e^+e^-$ angular observables at very low $q^2$. Although these observables are not sensitive to the violation of LFU, being dominated by the photon pole, they can provide important constraints on the Wilson coefficients $C_7^{(\prime)}$.} the $\chi^2$ deteriorates to $125.8$ for $91$ measurements, corresponding to $p=0.91\%$. The observables with the biggest individual tensions are listed in table~\ref{tab:tensions}.
It should be noted that the observables in this table are not independent. For instance, of the set $(S_5,F_L,P_5')$, only the first two are included in the fit as the last one can be expressed as a function of them\cite{Descotes-Genon:2013vna}$^{,\,}$\footnote{Including the last two instead leads to equivalent results since we include correlations as mentioned above; this has been checked explicitly.}.

\begin{table}[tb]
\renewcommand{\arraystretch}{1.5}
\centering
\begin{tabular}{ccccccc}
\hline
Decay & obs. & $q^2$ bin & SM pred. & \multicolumn{2}{c}{measurement} & pull\\
\hline
\input{obstable2s}
\hline
\end{tabular}
\caption{Observables where a single measurement deviates from the SM by $1.9\sigma$ or more (cf.~\protect\cite{Straub:2015ica} for the $B\to K^*\mu^+\mu^-$ predictions at low $q^2$).}
\label{tab:tensions}
\end{table}

\subsection{Implications for Wilson coefficients}

Next, we have performed fits where a single real Wilson coefficient at a time is allowed to float. The resulting best-fit values, 1 and $2\sigma$ ranges, pulls, and $p$ values are shown in table~\ref{tab:1Dbounds}. The best fit is obtained for new physics in $C_9$ only, corresponding to a $3.7\sigma$ pull from the SM. A slightly worse fit with a pull of $3.1\sigma$ is obtained in the $SU(2)_L$ invariant direction $C_9^\text{NP}=-C_{10}^\text{NP}$. This direction corresponds to an operator with left-handed leptons only and is predicted by several NP models. If we include $b\to se^+e^-$ observables in the fit and assume NP to only affect the $b\to s\mu^+\mu^-$ modes, the pulls of these two scenarios increase to $4.3\sigma$ and $3.9\sigma$, respectively.

\begin{table}[tb]
\renewcommand{\arraystretch}{1.5}
\centering
\begin{tabular}{cccccc}
\hline
Coeff. & best fit & $1\sigma$ & $2\sigma$ & $\sqrt{\chi^2_\text{b.f.}-\chi^2_\text{SM}}$ & $p$ [\%] \\
\hline
\input{1dbounds}
\hline
\end{tabular}
\caption{Constraints on individual Wilson coefficients, assuming them to be real, in the global fit to 88 $b\to s\mu^+\mu^-$ measurements. The $p$ values in the last column should be compared to the $p$ value of the SM, $2.1\%$.}
\label{tab:1Dbounds}
\end{table}

\begin{figure}[tb]
\includegraphics[width=0.45\textwidth]{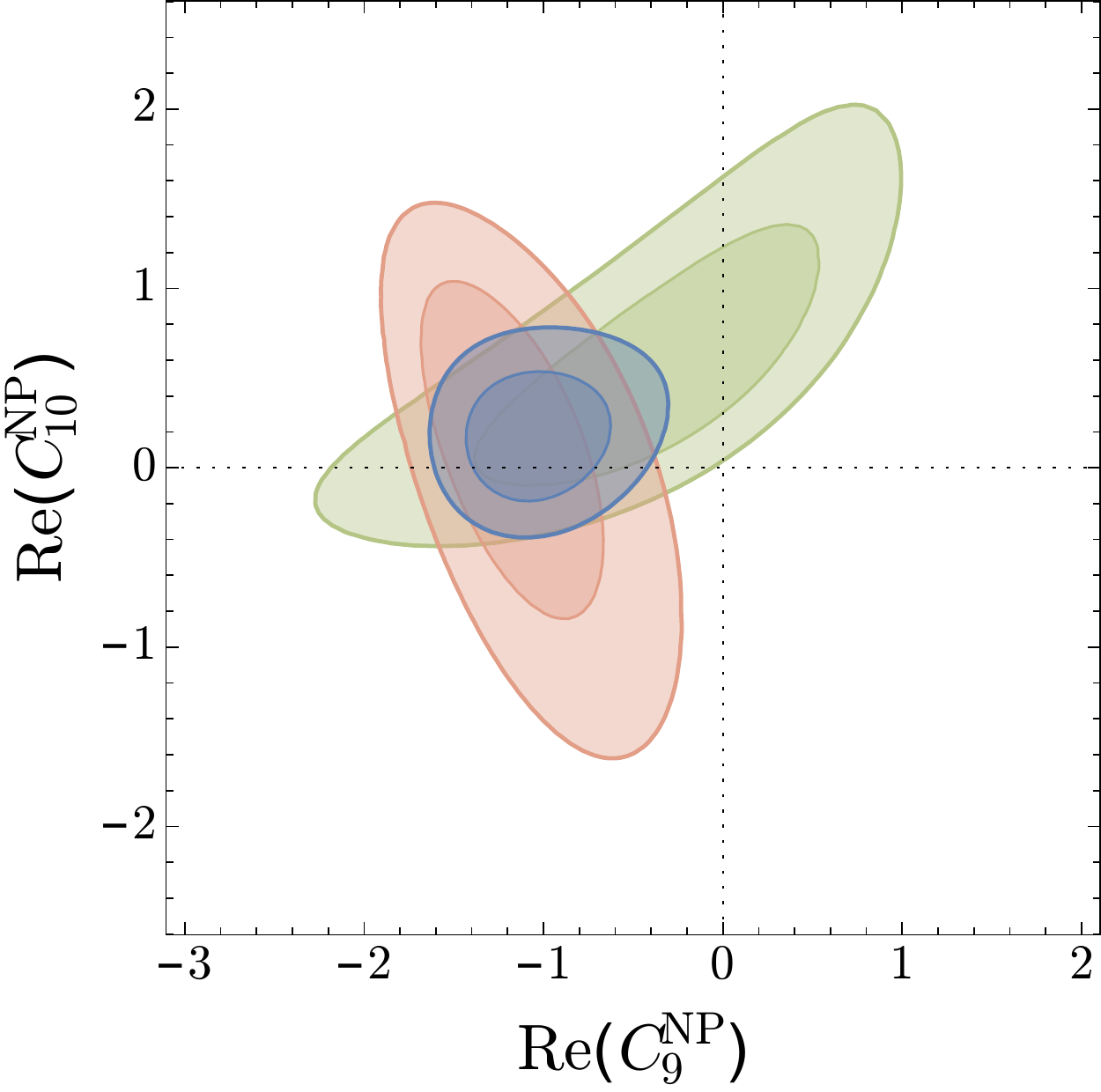}%
\hspace{0.05\textwidth}%
\includegraphics[width=0.45\textwidth]{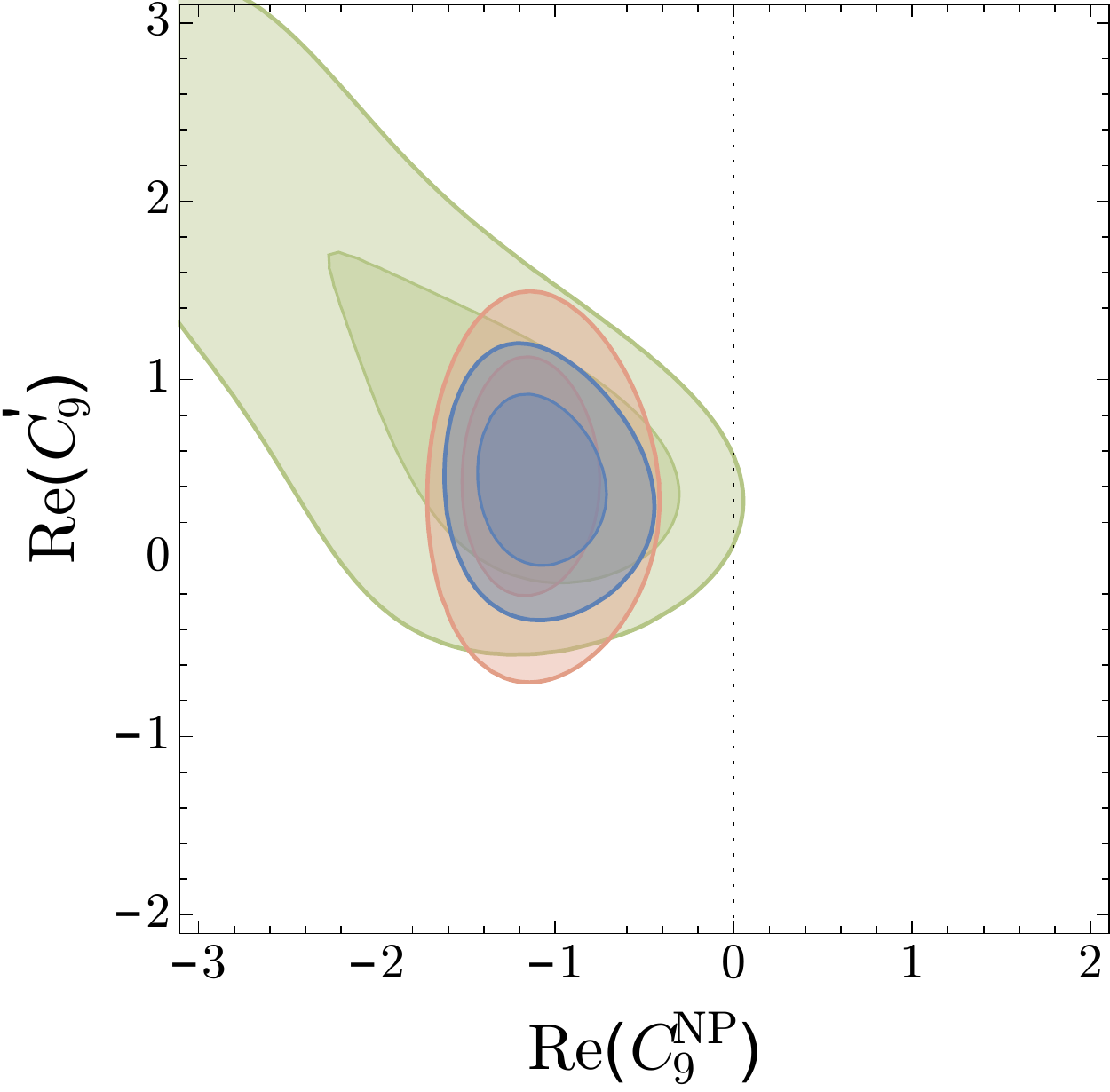}%
\caption{Allowed regions in the Re$(C_9^\text{NP})$-Re$(C_{10}^\text{NP})$ plane (left) and the Re$(C_9^\text{NP})$-Re$(C_9')$ plane (right). The blue contours correspond to the 1 and $2\sigma$ best fit regions from the global fit. The green and red contours correspond to the 1 and $2\sigma$ regions if only branching ratio data or only data on $B \to K^* \mu^+\mu^-$ angular observables is taken into account.}
\label{fig:obs}
\end{figure}

Allowing NP effects in two Wilson coefficients at the same time, one obtains the allowed regions shown in fig.~\ref{fig:obs} in the $C_9$-$C_{10}$ plane and the $C_9$-$C_9'$ plane. Apart from the $1\sigma$ and $2\sigma$ regions allowed by the global fit shown in blue, these plots also show the allowed regions when taking into account only $B\to K^*\mu^+\mu^-$ angular observables (red) or only branching ratio measurements of all decays considered (green).

\subsection{New physics vs. hadronic effects}

\begin{figure}
\centering
\includegraphics[width=0.6\textwidth]{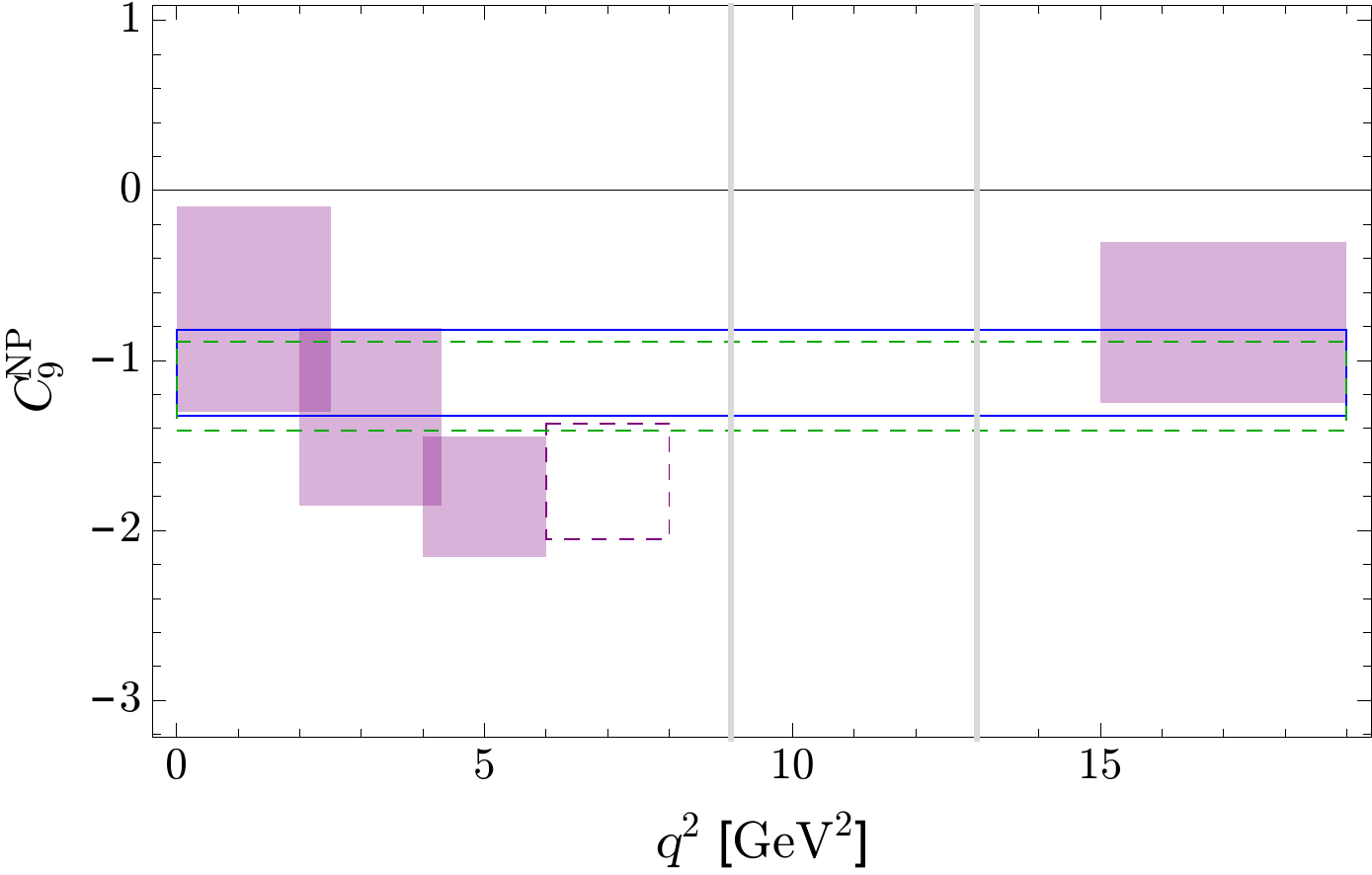}
\caption{Purple: ranges preferred at $1\sigma$ for a new physics contribution to $C_9$ from fits to all $B\to K^*\mu^+\mu^-$ observables in different bins of $q^2$. Blue: $1\sigma$ range for $C_9^\text{NP}$ from the global fit (cf.\ tab.~\ref{tab:1Dbounds}).
Green: $1\sigma$ range for $C_9^\text{NP}$ from a fit to $B\to K^*\mu^+\mu^-$ observables only.
The vertical gray lines indicate the location of the $J/\psi$ and $\psi'$ resonances, respectively.
}
\label{fig:WCq2}
\end{figure}

The result that the best fit is obtained by modifying the Wilson coefficient $C_9$ might be worrying as this is the coefficient of an operator with a left-handed quark FCNC and a vector-like coupling to leptons; non-factorizable hadronic effects are mediated by virtual photon exchange and thus also have a vector-like coupling to leptons (and the left-handedness of the FCNC transition is ensured by the SM weak interactions). It is therefore conceivable that unaccounted for hadronic effects could mimic a new physics effect in $C_9$. There are at least two ways to test this possibility.
\begin{enumerate}
 \item The hadronic effect cannot violate LFU, so if the violation of LFU in $R_K$ (or any of the other observables suggested, e.g., in~\cite{Altmannshofer:2014rta}) is confirmed, this hypothesis is refuted;
 \item There is no {\em a priori} reason to expect that a hadronic effect should have the same $q^2$ dependence as a shift in $C_9$ induced by NP.
\end{enumerate}
Let us focus on the second point. With the finer binning of the new LHCb $B\to K^*\mu^+\mu^-$ angular analysis, it is possible to determine the preferred range of a hypothetical NP contribution to $C_9$ in individual bins of $q^2$. To this end, we have splitted all measurements of $B\to K^*\mu^+\mu^-$ (including braching ratios and non-LHCb measurements) into sets with data below 2.3~GeV$^2$, between 2 and 4.3~GeV$^2$, between 4 and 6~GeV$^2$, and above 15~GeV$^2$ (the slight overlap of the bins, caused by changing binning conventions over time, is of no concern as correlations are treated consistently). The resulting $1\sigma$ regions are shown in fig.~\ref{fig:WCq2} (the fit for the region between 6 and 8~GeV$^2$ is shown for completeness as well but only as a dashed box because we assume non-perturbative charm effects to be out of control in this region and thus do not include this data in our global fit). We make some qualitative observations, noting that these will have to be made more robust by a 
dedicated numerical analysis.
\begin{itemize}
\item The NP hypothesis requires a $q^2$ independent shift in $C_9$. At roughly $1\sigma$, this hypothesis seems to be consistent with the data.
\item If the tensions with the data were due to errors in the form factor determinations, naively one should expect the deviations to dominate at one end of the kinematical range where one method of form factor calculation (lattice at high $q^2$ and LCSR at low $q^2$) dominates. Instead, if at all, the tensions seem to be more prominent at intermediate $q^2$ values where both complementary methods are near their domain of validity and in fact give consistent predictions\cite{Straub:2015ica}.
\item There does seem to be a systematic increase of the preferred range for $C_9$ at $q^2$ below the $J/\psi$ resonance, increasing as this resonance is approached. Qualitatively, this is the behaviour expected from non-factorizable charm loop contributions. However, the central value of this effect would have to be significantly larger than expected on the basis of existing estimates~\cite{Muheim:2008vu,Khodjamirian:2010vf,Jager:2012uw,Lyon:2014hpa,Jager:2014rwa}, as conjectured earlier \cite{Lyon:2014hpa}.
\end{itemize}
Concerning the last point, it is important to note that a charm loop effect does not have to modify the $H_-$ and $H_0$ helicity amplitudes\footnote{The modification of the $H_+$ amplitude is expected to be suppressed~\cite{Jager:2012uw,Jager:2014rwa}.} in the same way (as a shift in $C_9$ induced by NP would). Repeating the above exercise and allowing a $q^2$-dependent shift of $C_9$ only in one of these amplitudes, one finds that the resulting corrections would have to be huge and of the same sign. It thus seems that, if the tensions are due to a charm loop effect, this must contribute to both the $H_-$ and $H_0$ helicity amplitude with the same sign as a negative NP contribution to~$C_9$.

\section{Summary and Outlook}

The new LHCb measurement of angular observables in $B\to K^*\mu^+\mu^-$ is in significant tension with SM expectations. An explanation in terms of new physics is consistent with the data. Models with a negative shift of $C_9$ or with $C_9^\text{NP}=-C_{10}^\text{NP}<0$ give the best fit to the data. These findings are in very good agreement with preliminary results from a similar analysis presented at this conference~\cite{DHMVtalk}.

Arguments have been given why the tension being caused by underestimated form factor uncertainties, suggested~\cite{Jager:2014rwa} as an explanation of the original $B\to K^*\mu^+\mu^-$ anomaly~\cite{Aaij:2013qta}, does not seem to be supported by the data. A detailed numerical analysis of this point, with the help of the new LCSR result~\cite{Straub:2015ica} (and possibly the relations in the heavy quark limit~\cite{Jager:2012uw,Descotes-Genon:2014uoa,Jager:2014rwa} as a cross-check) would be interesting.

An important cross-check of the NP hypothesis is the $q^2$ dependence of the preferred shift in $C_9$ and it has been argued that also an unexpectedly large charm-loop contribution at low $q^2$ near the $J/\psi$ resonance could solve, or at least reduce, the observed tensions. A possible experimental strategy to resolve this ambiguity could contain, among others, the following steps.
\begin{itemize}
\item Testing LFU in the $B\to K^*\mu^+\mu^-$ vs.\ $B\to K^*e^+e^-$ branching ratios and angular observables, where spectacular deviations from the SM universality prediction would occur if the $R_K$ anomaly is due to NP~\cite{Altmannshofer:2014rta,Hiller:2014ula,Jager:2014rwa}, which can be accomodated in various NP models with a $Z'$ boson~\cite{Altmannshofer:2014cfa,Buras:2014fpa,Glashow:2014iga,Bhattacharya:2014wla,Crivellin:2015mga,Altmannshofer:2014rta,Crivellin:2015lwa}~\footnote{Some $Z'$ models~\cite{Gauld:2013qba,Buras:2013qja,Gauld:2013qja,Buras:2013dea} predict LFU to hold but could still solve the $B\to K^*\mu^+\mu^-$ anomaly.} or leptoquarks~\cite{Hiller:2014yaa,Biswas:2014gga,Buras:2014fpa,Gripaios:2014tna,Sahoo:2015wya,Varzielas:2015iva};
\item Searching for lepton flavour violating $B$ decays like $B\to K^{(*)} e^\pm\mu^\mp$, because in leptoquark models explaining the $B\to K^*\mu^+\mu^-$ anomaly, either $R_{K^{(*)}}$ deviates from one or lepton flavour is violated~\cite{Buras:2014fpa,Varzielas:2015iva} and also in $Z'$ models these decays could arise~\cite{Glashow:2014iga}.
\item Measuring the T-odd CP asymmetries\cite{Bobeth:2008ij,Altmannshofer:2008dz} $A_{7,8,9}$, which could be non-zero in the presence of new sources of CP violation.
\item Measuring BR($B_s\to\mu^+\mu^-$) more precisely as a clean(er) probe of $C_{10}$.
\end{itemize}
The first three items are null tests of the SM and could unambiguously prove the presence of new physics not spoiled by hadronic uncertainties; the last one is at least much cleaner than semi-leptonic decays.

On the theory side,
the new more precise data could be used, in the spirit of fig.~\ref{fig:WCq2}, to extract the preferred size, $q^2$ and helicity dependence of a possible hadronic effect, assuming the SM.
Combined with a better understanding of the charm-loop effect and more precise estimates of its possible size, this could shed light on the important question whether the effect observed by LHCb is the first evidence for physics beyond the Standard Model, or our understanding of strong interaction effects in rare semi-leptonic $B$ decays has to be revised. Both possibilities will have important implications.

\section*{Acknowledgments}

D.S.\ would like to thank the organizers of the 50th Rencontres de Moriond for the invitation, the members of the LHCb collaboration for sharing unpublished results on the $B\to K^*\mu^+\mu^-$ angular analysis, and Patrick Koppenburg, Christoph Langenbruch, Guy Wilkinson, and Joaquim Matias for useful discussions.
We thank Andreas Crivellin for valuable comments on the manuscript.
The research of D.S.\ is supported by the DFG cluster of excellence ``Origin and Structure of the Universe''.
Research at Perimeter Institute is supported by the Government of Canada through Industry Canada and by the Province of Ontario through the Ministry of Economic Development \& Innovation.

\bibliography{bsll}

\end{document}

%% file: obstable2s.tex
$\bar B^0\to\bar K^{*0}\mu^+\mu^-$ & $F_L$ & $[2,4.3]$ & $0.81 \pm 0.02$ & $0.26 \pm 0.19$ & ATLAS & $+2.9$\\
$\bar B^0\to\bar K^{*0}\mu^+\mu^-$ & $F_L$ & $[4,6]$ & $0.74 \pm 0.04$ & $0.61 \pm 0.06$ & LHCb & $+1.9$\\
$\bar B^0\to\bar K^{*0}\mu^+\mu^-$ & $S_5$ & $[4,6]$ & $-0.33 \pm 0.03$ & $-0.15 \pm 0.08$ & LHCb & $-2.2$\\
$\bar B^0\to\bar K^{*0}\mu^+\mu^-$ & $P_5'$ & $[1.1,6]$ & $-0.44 \pm 0.08$ & $-0.05 \pm 0.11$ & LHCb & $-2.9$\\
$\bar B^0\to\bar K^{*0}\mu^+\mu^-$ & $P_5'$ & $[4,6]$ & $-0.77 \pm 0.06$ & $-0.30 \pm 0.16$ & LHCb & $-2.8$\\
$B^-\to K^{*-}\mu^+\mu^-$ & $10^{7}~\frac{d\text{BR}}{dq^2}$ & $[4,6]$ & $0.54 \pm 0.08$ & $0.26 \pm 0.10$ & LHCb & $+2.1$\\
$\bar B^0\to\bar K^{0}\mu^+\mu^-$ & $10^{8}~\frac{d\text{BR}}{dq^2}$ & $[0.1,2]$ & $2.71 \pm 0.50$ & $1.26 \pm 0.56$ & LHCb & $+1.9$\\
$\bar B^0\to\bar K^{0}\mu^+\mu^-$ & $10^{8}~\frac{d\text{BR}}{dq^2}$ & $[16,23]$ & $0.93 \pm 0.12$ & $0.37 \pm 0.22$ & CDF & $+2.2$\\
$B_s\to\phi\mu^+\mu^-$ & $10^{7}~\frac{d\text{BR}}{dq^2}$ & $[1,6]$ & $0.48 \pm 0.06$ & $0.23 \pm 0.05$ & LHCb & $+3.1$\\

%% file: 1dbounds.tex
$C_7^\text{NP}$ & $-0.04$ & $[-0.07,-0.01]$ & $[-0.10,0.02]$ & $1.42$ & $2.4$ \\ 
 $C_7'$ & $0.01$ & $[-0.04,0.07]$ & $[-0.10,0.12]$ & $0.24$ & $1.8$ \\ 
 $C_9^\text{NP}$ & $-1.07$ & $[-1.32,-0.81]$ & $[-1.54,-0.53]$ & $3.70$ & $11.3$ \\ 
 $C_9'$ & $0.21$ & $[-0.04,0.46]$ & $[-0.29,0.70]$ & $0.84$ & $2.0$ \\ 
 $C_{10}^\text{NP}$ & $0.50$ & $[0.24,0.78]$ & $[-0.01,1.08]$ & $1.97$ & $3.2$ \\ 
 $C_{10}'$ & $-0.16$ & $[-0.34,0.02]$ & $[-0.52,0.21]$ & $0.87$ & $2.0$ \\ 
 $C_9^\text{NP}=C_{10}^\text{NP}$ & $-0.22$ & $[-0.44,0.03]$ & $[-0.64,0.33]$ & $0.89$ & $2.0$ \\ 
 $C_9^\text{NP}=-C_{10}^\text{NP}$ & $-0.53$ & $[-0.71,-0.35]$ & $[-0.91,-0.18]$ & $3.13$ & $7.1$ \\ 
 $C_9'=C_{10}'$ & $-0.10$ & $[-0.36,0.17]$ & $[-0.64,0.43]$ & $0.36$ & $1.8$ \\ 
 $C_9'=-C_{10}'$ & $0.11$ & $[-0.01,0.22]$ & $[-0.12,0.33]$ & $0.93$ & $2.0$ \\ 